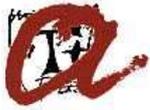

**ETSEQ, Tarragona**
**FMCS, Reus**
**Universitat Rovira i Virgili**

# Hepatocyte Aggregates:

## *METHODS OF PREPARATION IN THE MICROGRAVITY SIMULATING BIOREACTOR USE IN TISSUE ENGINEERING*

**Research work submitted for Diploma d'Estudis Avançats degree**

Presented by: Verónica Saravia
Supervised by: Dr. Petros Lenas
Dr. José Luis Toca

**Tarragona, September 2005**

The supervisor Dr. José Luis Toca-Herrera, has revised and approved for submission the Master Dissertation, the research work title: "Hepatocyte aggregates: Methods of preparation in the microgravity simulating bioreactor , use in tissue engineering", presented by Verónica Saravia.

Dr. José Luis Toca-Herrera, Universitat Rovira i Virgili
Tarragona.

## Contents

# Figure index

# Abstract

Tissue Engineering concerns the three-dimensional cell growth so that bio-artificial tissues could be created and used for transplantation. Looking at the history of Tissue Engineering we could see that very few of the promises have been implemented in clinics. The recently expressed concerns from the Tissue Engineering research community for a re-direction of the research activities necessitate the proposition of new methodologies. The methodology we propose here in general terms has to do with the simulation in bioreactor systems of liver structures as are described in liver anatomy so that the hepatocyte microenvironments that determine their function could be re-created in vitro.

Much of the time for this project has been devoted in preliminary design of this approach, reviewing information from literature for the phenomena involved in the establishment of liver tissue functions from cooperative cell interactions and making the most methodological and compatible choices of the bioreactor systems.

The approach needs the use of hepatocyte aggregates as entities to load the bioreactor systems. A new bioreactor, the microgravity simulating rotation bioreactor, has been used for the preparation of cell aggregates. Two methods have been applied, the aggregation of cells alone and the aggregation with the involvement of microcarriers (Cytodex 3) on which the cells attach.

HepG2 cell aggregates with microcarriers and spheroids were prepared, setting the conditions for the production, protocols to monitor the cell culture, and their characterization. Direct count with trypan blue and qualitative image analyses were used to characterize the aggregates. Preliminary toxicology assays with acetaminophen (APAP) as toxicant in a range of 5 to 80 mM were carried out using standard culture and the aggregates. Total protein, DNA content and viability assayed with MTT were used to evaluate the toxicity. As a consequence of the results obtained, improvements in the methodology are needed. New experimental techniques, such as the assessment of cytochrome P450 isozyme activity will be used to determine the feasibility of the 3 D model (in comparison with the standard 2D culture).

In order to be able to study the heterogeneity within aggregate confocal microscopy is proposed as a valid method to incorporate in the characterization of the aggregates as well as in the toxicological tests.

Microcontact printing was used successfully to produce a patterned surface. This surface was tested for BSA protein and the presence of the pattern confirmed by fluorescence microscopy and AFM. Immobilization of cell aggregates on protein-patterned surfaces will be attempt.

Once ready to move to the next stage, the work will consist in the production of primary rat hepatocytes aggregates following the methods developed and it use in bioreactors with different configurations aiming to gain further understanding of the role of cell heterogeneity in the cooperative behaviour of cells in vitro.

# 1. Introduction

Tissue engineering is a field of few decades history and involves several different areas such as cell and molecular biology, materials science, reactor engineering, and clinical research [1]. As it was defined in a workshop of the National Science Foundation of USA in 1993, it involves the *"the application of principles and methods of engineering and life sciences toward fundamental understanding of structure-function relationships in normal and pathological mammalian tissues and the development of biological substitutes to restore, maintain or improve tissue function"*.

## *1.1 The significance of the three-dimensional cell growth*

Tissue Engineering initiated as a distinct research field when the concept of cell growth in three-dimension (3D) was introduced. This was implemented by Bob Langer, a Chemical Engineer in MIT and Joseph Vacanti, a Medical Doctor in Harvard in 1980 [2], seeding cells in porous biomaterials and suspending them in medium containing nutrients, oxygen, and growth factors in a bioreactor system. These scaffolds, whose pore network simulates the *in vivo* vascular network, make possible the maintenance of the cell viability with the adequate provision of nutrients and oxygen inside the cell mass. The cells spread, move, divide and occupy the external surface of the material.

The three-dimensional cell growth does not only resembles the geometry of real tissues, it is rather a basic requirement for the cells to retain their differentiated phenotype through cell-to-cell interactions. It has been proven that cells taken from tissues and placed in two-dimensions on a culture dish, gradually loose the tissue functions that were performing, and soon die. In contrast, cells in aggregates are more functional and live longer. This is a general trend observed for many cell types, e.g., hepatocytes [3], or chondrocytes [4].

## *1.2 The current situation in Tissue Engineering*

Why *in vitro* tissue formation cannot be easily optimized

The research efforts in Tissue Engineering have concentrated until now in the design of in vitro systems (i.e. scaffolds, bioreactors, media and differentiation factors) improving cell level indices (e.g. longevity of cell, viability and differentiation markers). Considerable progress was made at the technical level if we take into account that few decades ago it was impossible to keep the cells taken from organisms viable *in vitro* for the time needed for tissue development. Unfortunately however, none of these indices is directly related with the tissue function, it is rather their combination that determines the tissue formation and function. What is the appropriate combination of cell characteristics, and how this combination arises during the in vitro culture from an optimal combination of the design parameters such as cell types, material scaffold properties, growth factors and bioreactor systems, has not been found yet making bio-artificial tissues sub-optimal in their functionality. The interrelation of the design parameters and the unknown way they influence cellular functions make impossible a methodological approach as in other Engineering fields (e.g., an increase in the porosity of the scaffold may increase the cell viability due to better oxygenation, but may also prevent differentiated functions with the dilution of growth factors that are secreted by the cells). Trial-and-error methods remain the only approach, having however a high cost as it is reflected in the high risk of product development of the Tissue Engineering companies. Re-engineering a candidate product to achieve better properties proved to be a very difficult problem [5]. Some of the recent failures of the private sector as the bankruptcy of Advanced Tissue Science and Organogenesis, were attributed to the lack of basic understanding [6].

Consequently until now, the successful developments of tissue engineering are just a few. The skin grafts, which were one of the first therapies to be approved, are commercially available. Cartilage is also being used and it is in clinical trial. Although researchers are currently pursuing tissue-engineered structures for a host of tissues including bone, liver, artery, nerve, pancreas, skin, kidney, and bladder, we have to wait several decades before bio-artificial tissues will be available for transplantations.

According to Langer and Vacanti pioneers of Tissue Engineering *"in the next three decades, medical science will move beyond the practice of transplantation and into the era of fabrication"*.

The improvements that have been made in the three-dimensional cell growth provide opportunities for several other applications such in *in vitro* toxicity tests, *in vitro* tests for drugs, or *in vitro* basic research studies of tissue physiology. Even though, bio-artificial tissues developed until now are far from the real tissues in functionality, they are closer to real tissues than the cells grown on two-dimensional dishes, and then represent a better model for study.

## *1.3 Bio-artificial liver devices*

Bio-artificial tissues although not optimal, are an option in the cases where no alternative method to cure a life threatening disease is available. Over 30,000 patients die every year in US from acute liver failure [7]. Although liver transplantation can save these patients, no available livers can be found in time. Unfortunately 1000 patients (80% of those with acute liver failure), die every year in US while awaiting a donor liver for transplantation [8]. A temporal support of the patient with a liver support device until a liver is found for transplantation is therefore considered the only alternative solution. In addition, the acute liver failure is reversible in some cases if the liver functions are substituted temporally until the liver of the patient recovers.

There are several designs of bio-artificial liver devices, which reflect the multiple criteria that must be considered, as transport of nutrients and oxygen, immunoisolation, minimum amount of liver cells required to replace the liver cells, etc. [7], [9], [10]). The different designs include hollow fiber, flat plate, perfused beads/scaffolds bioreactors, etc. These bioreactors have two compartments, one for the hepatocytes and the other for the blood of the patient, e.g. hepatocytes are outside the hollow fibers through which the blood of the patient is perfused, (Fig. 1). In most of the cases the hepatocytes are forming cell aggregates, attached on scaffolds or encapsulated in polymers, so that their differentiated functions could be kept for long time in vitro.

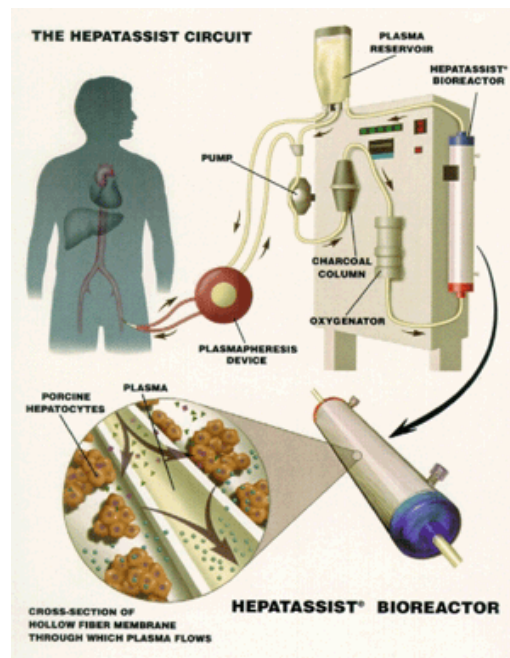

*Figure 1: The HepatAssist bio-artificial liver device developed by Circe Biomedical of Lexinghton, Massachesetts based on the work of Prof. Demetriou, (see text), and used in clinical trials in Spain, Reina Sofia Hospital, Cordoba and Clinico de Barcelona. The blood of the patient is perfused through the fibers of a hollow fiber bioreactor and it is cleaned by hepatocytes residing in the space between the fibers.*

Several companies are in clinical trials, some examples are the Vitagen, the Hepatix, the Circe Biomedical, the Algenix, the Excorp Medical, and the Charite Virchow Clinic-Berlin [7]. The first case of a patient treated in Spain, at the Unidad de Cuidados Intensivos, Instituto de Enfermedades Digestivas, Servicio de Hepatologia, Barcelona, was reported in 2001 [11].

Clinical trials until now have proven the safety of various bio-artificial liver devices. However, no evidence of the efficacy in treating the acute liver failure could be found yet, even with the most widely used bio-artificial liver device, HepatAssist, whose development has taken several years ([12], [13]). The most probable reasons for this failure are the unknown etiology of the acute liver failure in most of the cases, the variability of the symptoms caused by the disease, and the inability to design functional liver tissue in the devices or to improve their functions when needed [14]. The use of sliced liver tissues, which present the original liver structure, work better than any other hepatocyte three-dimensional arrangement made in vitro [15]. However, sliced liver tissue cannot solve the problem due to the short life span of such cultures cause by the inappropriate cell nutrition, and the limiting number of hepatocytes, that cannot multiply easily in the amount needed.

Further development in bio-artificial liver devices remains as an open question. None of the design parameters of these devices refers to liver tissue functions, so a methodological optimization could be tried. All the design parameters influence the tissue functions but indirectly, in a combined and unknown way. The uncertainty for the next step in Tissue Engineering has been recognized officially. *"Further basic science studies in tissue engineering systems"* has been recommended at the Symposium of Reparative Medicine in National Institute of Health (NIH) of US, together with the *"development of a vision for Tissue Engineering"*, and a replacement of trial-and-error methods with *"rational design approaches"*, but no recommendation has been provided on how the above changes could be accomplished [16].

## 2. Hypothesis

*A new conceptual framework for the definition of objectives*

Two conclusions can be drawn from the history of successes and failures in Tissue Engineering. First, the three-dimensional cell growth is an almost absolute requirement for the retention of cell viability and functionality. Therefore in order to have the experimental basis for any "basic science study" in Tissue Engineering as recommended by NIH, a well-controlled method for three-dimensional cell growth is needed irrespectively of the type of studies that will follow.
The other conclusion is that the three-dimensional cell growth in the bioreactor tried until now is not the only requirement to restore tissue functions in bio-artificial tissues. How can this happen? If cell functionality is maintained in three-dimensional cell growth why the bio-artificial tissue is not functional? Is it a problem that could be solved optimizing the functionality of the cells with further trial-and-error? Or is a problem with the "basic understanding" of how a bio-artificial tissue should be designed?

In view of the official recognition in the Symposium of NIH mentioned above for a new vision in Tissue Engineering, it makes no sense to propose various combinations of bioreactors, scaffolds, cell types and growth factors adding another unsuccessful combination from the numerous possible. It should be rather proposed a new vision and check experimentally if it can be methodologically implemented.

We think that a underestimated until now in Tissue Engineering aspect of tissues, that of the spatio-temporal organization of the cell activities to tissue functions, deserves further consideration as a candidate concept on which new research directions could be developed. A tissue is not a random three-dimensional collection of cells according to anatomical and physiological data. The cells are organized in space so that they can properly cooperate. It is the cell cooperation that integrates over the whole space of the tissue that appears finally as tissue functions.

The above is not a theoretical consideration, it is what tissue physiology have proved experimentally *in vivo* and *in vitro* for various tissues. For example the regulation of the concentration of glucose in the blood is not a hepatocyte function but a liver function, i.e., it cannot be performed by a single hepatocyte or by a random three-dimensional arrangement of hepatocytes in scaffolds. This function appears only by the cooperation of metabolically heterogeneous hepatocytes that have complementary metabolic functions, one that secretes glucose and one that consumes glucose (see Annex 1 for more details). These two types of hepatocytes activate their metabolic functions at different times according to the level of glucose in the blood [17]. Similarly the control of the cartilage growth during its development cannot be achieved by a single chondrocyte or by a random three-dimensional collection of chondrocytes in scaffolds. The control function it is not a cell function but a tissue function build by the cooperation of different chondrocyte types that exchange protein signals forming a negative feedback loop among chondrocytes in early and late differentiation stages that keeps the differentiation rate inside some limits [18]. Another example is the controlled release of insulin, which is a property of pancreatic islets and not a property of beta-cells [19].

In conclusion two basic properties build the real tissues, the cell heterogeneity concerning the cell functions and the cell cooperativity concerning the complementarity of these functions. As far as we know, it has not been checked in bio-artificial tissues if such functional cell heterogeneity has appeared, although numerous data exist in tissue physiology for the significance of its role. A reason for the focus of until now Tissue Engineering studies on the cell instead of the tissue level, could be the choice of the cell and the levels below the cell (i.e., genes, protein) as a reference system. But it seems that this has started to change, and Biology moves now from the cell and gene level to the levels above. For example research priorities in cancer have shifted from the genes to the interactions among cancer cells and other cells in the tumor, treating the tumor as a pathological tissue, (among the new research priorities defined in 2003 by the Cancer Research Institute of US is the "characterization of the interaction among a tumor, its microenvironment and the entire body" and the "definition of the dynamic communication among cancer cells and surrounding cells"). Therefore, this is a general re-direction in biomedical research and Tissue Engineering cannot remain outside.

## *Design of culture systems where hepatocytes exhibit functional heterogeneity*

The implementation of the concepts mentioned above of cell heterogeneity and cooperatively, however is not straightforward. A major difficulty addressing this problem experimentally arises from the complexity in the establishment of cell functions, i.e., cell functions are determined by the cell microenvironment but it is the cells by themselves that determine this microenvironment. For example, the hepatocytes that either release or consume glucose, they do so according to the oxygen concentration in their microenvironment [20], which however is determined besides the fluidics of the system by the hepatocytes themselves that consume the oxygen with different and not yet determined way.

To face this difficulty we propose to transfer information from liver physiology and anatomy to gradually more complicated bioreactor designs that resemble the liver structure. We will start with the preparation of hepatocyte aggregates but we do not consider these aggregates as components of a bio-artificial liver as it was until now. We consider them rather as components of another bioreactor that resembles the basic structure in the liver, the liver sinusoid, (see Annex 1), where we will try to implement the hepatocyte heterogeneity. We further consider the bioreactor-sinusoid as a component of another bioreactor system composed of several sinusoids, where we will try to implement the hepatocyte cooperativity. In addition to the use of conventional bioreactors such as tubular or packed-bed, for the establishment of tissue functions, we will use micro-patterned surfaces or combination of surfaces where cells or cell aggregates are placed accurately, to simulate in smaller scale the liver structures.

The culture systems proposed here are not bio-artificial liver devices. They are rather experimental model-systems from which new design criteria referring to the tissue level for bio-artificial liver could be determined. These experimental model-systems that are at the moment designed using qualitative information when functional could be used for the extraction of mathematical formulas that describe quantitatively the way heterogeneous hepatocytes interact cooperatively, offering quantitative design criteria. For a bio-artificial liver to be developed these criteria should be combined with several other that are related with the characteristics of the acute liver failure, e.g., toxic compounds in the blood of patients, number of hepatocytes needed for quick cleaning of the blood, etc., which is far from the present project.

In our case we start with cell aggregates with a concrete plan for a methodological upgrading of the structural complexity of the in vitro system (based on liver physiology and anatomy), so that chances to move from the heterogeneity to cooperativity will be increased.

## *Basic unit: Cell aggregates*

Cells self-organize in vitro to form multicellular structures under suitable conditions. There are several methods to prepare cell aggregates. Roughly they can be divided in two groups [21]: i) stationary culture using for example dishes not covered with collagen (for anchorage-dependent cells the biological dishes are covered with collagen to facilitate the cell attachment) or dishes covered with agar, where the cells are prevented to attach on the surface, and ii) non-stationary culture such as shaking or spinning culture using spinner flasks. These methods are widely used, however, they have the problem that imposes severe physical constraints, so that the cells should have a significant tendency to aggregate in order to attach one to another and form aggregates. For example in the shaking cultures, the shaking needed to keep the cells in suspension makes the cells to move fast, then they have no time to develop their connections. In addition the shear-stress is harmful to cell membranes.

The first important step is the preparation of cell aggregates choosing a suitable bioreactor system that can give control on the aggregate properties and optimizing the conditions for the formation of aggregates.

The Engineers of NASA designed a bioreactor that simulates microgravity (Fig. 2) in order to have a system on earth for comparison with their experiments in space, [22].

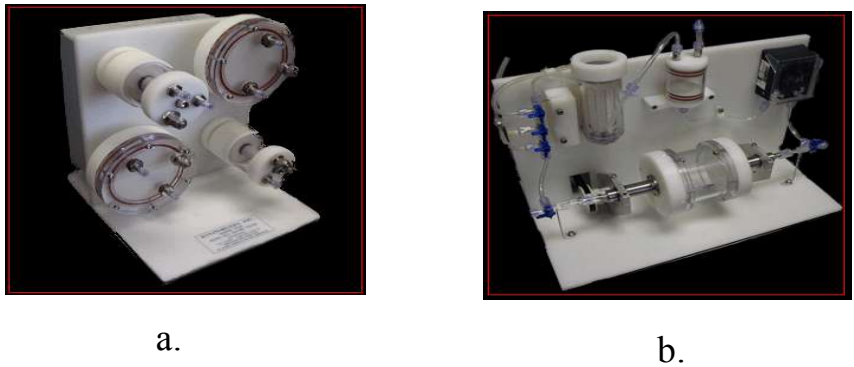

a.  b.

*Figure 2: The various types of rotation bioreactors. a. Batch type rotating vessel, (used in this studies), and rotating wall vessel, (cylinder). The second is used for large cell-biomaterial constructs. b. Perfused type used for long-term cultures.*

The "simulated microgravity" environment of the rotating bioreactor combines several beneficial factors for cell culture in tissue engineering. First, fluid flows in a laminar regime at most operating conditions, avoids the large shear stresses associated with turbulent flow and allows introduction of controlled and nearly homogenous shear fields. The suspended cells rotate as a solid body, maintaining their relative positions for long periods, allowing them to touch one another and find the appropriate receptors for connection on their surface, or to construct cell-bridges between cells growing on microcarriers beads. The low-shear environment, in combination with randomized gravitational vectors, restricts the diffusion of factors secreted by the cells and needed for cell growth, away from the cell microenvironment.

It was proven that under the conditions provided in this bioreactor the cells aggregate easily forming the so-called spheroids, without necrotic cores, which mean that transport of nutrients was adequate without turbulent flow. While several studies have been performed after that time in Tissue Engineering using this bioreactor, (reference above), very few studies refer to hepatocytes, with first one that of Khaoustov et al., [23] where it was found superior performance in comparison with other systems in respect of the aggregate size and culture longevity.

**Microcarriers**

Cells grown on microcarriers in the rotation bioreactor will be used in this study in order to increase the rate of cell growth and size of the aggregates. Microcarriers are beads of polymer, on the surface of which cells can attach and grow. The first microcarriers from cross-linked dextran were developed in 1967 by van Wezel for the culture of anchorage depended cells in suspension [24]. Microcarriers facilitate the formation of large aggregates, that was shown for the first time in a study by Avgerinos [25]. Cells were cultivated in conventional microcarriers, (Cytodex 2, Pharmacia), and grew normally. After the cells reached confluence, the cells began to migrate toward one end of the microcarriers and form clumps that jointed the microcarriers, (Fig. 3). The growth rate of the cells is increased with microcarriers because anchorage-dependent cells do not grow fast if not attached. The increase in the growth rate makes faster the formation of large aggregates.

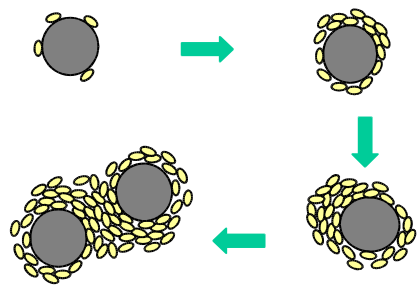

*Figure 3: Cells attached on microcarriers multiply and cover the surface in monolayers. Further cell growth results in multiple layers at several points that act as bridges between cell-populated microcarriers.*

# Objectives

**Objective 1 (short term)**

a)  **Preparation of HepG2 aggregates in the microgravity-simulating bioreactor with single cells and microcarriers**

An immortalized cell line, HepG2 is used considering that at this stage the work is mostly related with technical aspects of the culture systems. This will facilitate the experiments since these cells are readily available and grow indefinitely in culture.

b)  **Characterization of the aggregates**

c)  **Test of micro-patterning method**

The purpose at this stage is to test the patterning of surfaces with a protein for future application.

d)  **Preliminary toxicity test with hepatocyte aggregates**

In vitro hepatocyte systems are widely used in toxicity tests of drugs since the liver is the organ for the detoxification of compounds entering the blood. Since two different types of cell aggregates are formed in this project, an immediate use of them is toxicity tests. At this stage the purpose is the design of the experimental set-up for the toxicity tests, which includes the modification of the analyses usually performed on the cells in the commonly used tests with hepatocytes on flat surfaces.

**Objective 2 (Medium term)**

a.  Use of aggregates in toxicity tests and comparison with classical methods.
b.  Preparation of hepatocyte aggregates from primary hepatocyte

Although HepG2 cell lines are used in bio-artificial liver because they retain some of the differentiated properties of the *in vivo* hepatocytes, they are not suitable for studies related with the glucose metabolism. Glucose metabolism is altered in the immortalized cells making them to consume much more glucose than the *in vivo* cells, and it seems difficult to establish conditions were these cells could also release glucose, as it is required to establish metabolic cooperativity *in vitro*. Therefore the use rat primary cells, i.e., cells extracted from rat livers will be required. Immortalized cell lines may be adequate for toxicity tests, [26].

**Objective 3 (Long term)**

**Use of aggregates of primary hepatocytes in organotypic culture systems, conventional bioreactors and micro-patterned surfaces, specifically designed to resemble the liver structures**

# 3. Materials and Methods

## *3.1 Cells*

Human Hepatoma cell line HepG2 was used [27]. HepG2 passage nº 8 of ACAC nº HB-8065 was kindly proportioned by the Biochemistry department of Medical and Health Science School, Rovira i Rovira University in Reus, Tarragona. The cells were frozen, suspended in cell freeze media, and stored in liquid nitrogen tank until used.

## *3.2 Chemicals*

The basal medium used is DMEM (GIBCO), which concisely consist of glucose, essential aminoacids, vitamins and salts. It is complemented with 10% fetal bovine serum (FBS, GIBCO), which incorporate growth factors, lipids, hormones and minerals; 1% non-essential amino-acids 100X (GIBCO) which are glycine, L-alanine, L-asparagine, L-aspartic acid, L-glutamic acid, L- proline and serine; 1% penicillin-streptomycin 100X (GIBCO) and a 2,5% of 0,16 M solution of L-glutamine (SIGMA).
Other products used were trypsin–EDTA 10X (0.5% Trypsin, 5.3 mM EDTA•4Na, GIBCO), acetaminofen (APAP, from SIGMA), Hoechst 33344 (SIGMA), 3-[4,5-dimetiltiazol-2-il]2,5-difenil-tetrazolium brimide (MTT, SIGMA), and dimetil sulfoxide (DMSO, SIGMA).

## *3.3 Culture conditions*

The cell culture was carried out in a $CO_2$ incubator at 37ºC, 5% of $CO_2$ in air, and in a humidified atmosphere. Refer to annex B for more detail of the protocols used.

## *3.4 Measurements*

Quantification of the amount of cellular material was done by direct count in hemacytometer and measuring protein and DNA content. The total amount of protein was assayed by Biuret method in a Cobas Fara Mira analyzer. DNA was estimated measuring the fluorescence of Hoechst 33344 with an excitation filter of 355 nm and an emission filter of 460 nm, in a Fluoroskan spectrofluorimeter. It is based in the fact that Hoechst dye inserts in the DNA among rich regions of adenine-thymine, giving fluorescence that is proportional to the DNA content of the sample.
Glucose content was assayed using an enzymatic colorimetic test (GOD-PAP method from QCA S.A. Amposta, Spain) measuring at 546 nm with a Perkin Elmer Lambda 2 UV/VIS spectrophotometer.

The released of lactate deshydrogenase (LDH) was estimated using a kit based on SFBC technique from QCA S.A., Spain, in a Cobas Fara Mira analyzer. Since the LDH is a cytosol enzyme, the increase of LDH activity in the medium is an indicative of cell lysis.

 The viability of the cells was assayed using MTT. This method is base in the color change that MTT tetrazolim salt undergoes from yellow to dark insoluble formazan caused by the activity of various dehydrogenase enzymes in the cells. The tetrazolim ring is cleaved in active mitochondria, and so the reaction occurs only in living cells. The viability is expressed as a % of the viability of the cells in the control assay.

Before imaging the cells, to have a better visualization of them, MTT was added to stain the cells. In contrast to trypan blue dye exclusion, hepatocytes dyed by MTT are both viable and metabolically intact. Dead cells do not stain whereas dying cells may stain feebly and contain a few dark particles. The images were taken using a Reichert-Jung Polyvar microscope and a Nikon camera model.

## *3.5 Growth curve in 2D*

Tissue culture plates were seeded with $2 \times 10^5$ cell/ml per plate and 3 ml of complete medium. The cell growth was followed for 96 hours. Each day 3 plates were used for sampling. Samples of the medium and cells to assay the remaining glucose, LDH release and protein were taken. Each measurement was done in triplicate.

## *3.6 Growth in 3D*

*Bioreactor*

Disposable 10 ml bioreactors (Synthecom, Inc. Houston, TX) were bought to Cellon S.A. (Luxemburg). Before use they were rinsed with PBS and allowed to incubate over night. After a final rinse with complete medium, they were ready to use. For the cell culture the reactor was place inside the $CO_2$ incubator at 37ºC and the rotation adjusted. The sampling and the media exchange were done through ports that allow the extraction of medium maintaining aseptic conditions.

*Microcarriers*

Cytodex 3, collagen coated dextran beads of 175 μm diameter bought to Amersham Biosciences (Spain), were used as microcarriers. They were prepared and sterilized according to the manufacturer's instructions. Briefly, 0,03 gram of dry microcarrier beads were incubated with 1,5-3 ml of PBS at room temperature for at least 3 hours to allowed the beads to swell. After they were rinsed with PBS, replaced with new one, and sterilized by autoclaving at 115ºC, 15 minutes, 15 psi. Prior to use, sterilized microcarriers were allowed to settle, the supernatant was removed and microcarriers rinsed with warm 37ºC medium. Microcarriers were then ready to use for culturing. Cytodex that had been hydrated and sterilized was store in sterile PBS at 4ºC until needed.

*Cells*

Cells in a 75cm$^2$ flask in exponential growth were tryspinised and then used to form aggregates with or without microcarriers. The initialization of the culture is explained in the results section.

*Toxicological tests*

Toxicological tests were carried out with acetaminophen (APAP), a commonly use anti-inflammatory drug that can produce hepatotoxicity both *in vivo* and *in vitro*. The experimental design consisted in testing different concentrations of toxins (from 0,5mM to 50 mM) and its control, each of them in triplicate. Different concentration of toxicant is added to the culture medium of plate with cells in exponential growth phase (in 2D or 3D). In controls no additon of toxic was done. After 24 hours of the addition of the APAP cell mass estimation and viability test were carried out.

# 4. Results

## *4.1 Standard 2-D culture of HepG2*

The standard culture in two dimensions of the HepG2 cells was carried out in order to verify the correct growth of the cells in the medium chosen. The medium chosen enabled the cells to grow. As it is shown in figure 4 in the first 24 hours there was a lag phase where the cells recovered from the subculturing process and adapted to the new conditions. Between 24 and 72 hours, there was a growth period shown by the increase amount of biomass measured as total protein, with a decrease of glucose concentration. After 72 hours of culture, the growth rate of the culture decreased because of a combination of factors such as: glucose and other nutrients were in low amount or exhausted, accumulation of toxic metabolic byproducts such as lactate and ammonia and restriction of the available surface area.

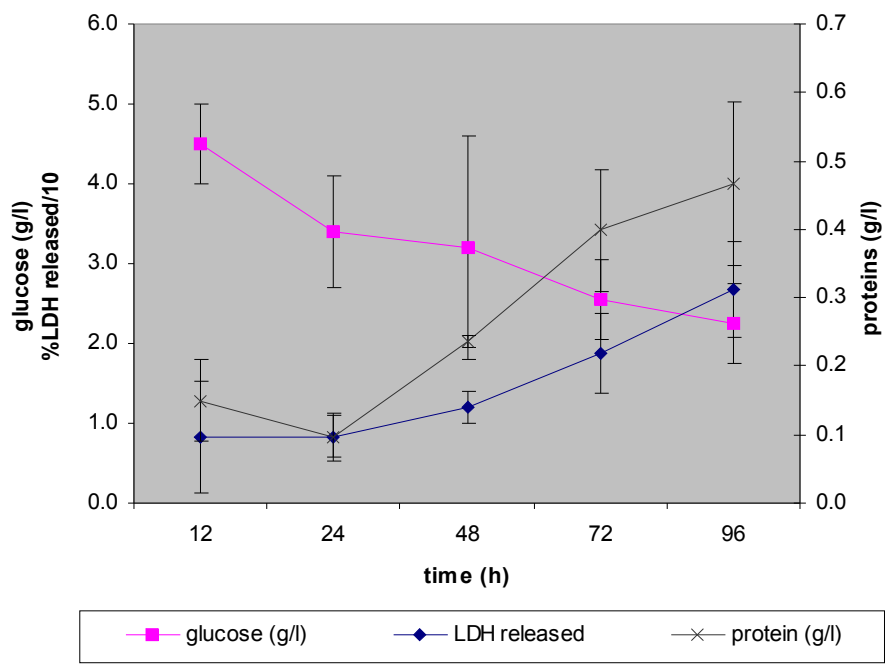

*Figure 4: Growth curves of standard culture of HepG2*

## *4.2 Bioreactor experiments*

The experiments were performed in the rotation bioreactor. Each experiment lasted at least a couple of weeks considering the time that the cells require to grow. Due to the availability of only one bioreactor only one experiment could be performed each time.

### *4.2.1 Culture in 3D: Cytodex*

Culture on Cytodex 3 enables the cells: i) to attach on the surface of beads, in this case spheres covered with denaturalized collagen, and ii) to grow the cells in the suspension culture mode, in the bioreactor
The settings of the culture had to be fixed, that means: initiation of the culture, speed of rotation and medium exchange.
First, the initiation of the culture was optimized aiming to obtain the coverage of the surface of the Cytodex and minimize the number of cells in suspension not attached to micro-carriers. Taking into account the recommendations of Cytodex manufacturers about the amount of microcarriers to be use and size of inoculum, several experiences were carried out. An inoculum of $4 \times 10^6$ cells and 0,03 g Cytodex were directly placed in the bioreactor. It was filled with complete medium and rotation adjusted to 8 rpm. The attachment of the cells was monitored by observation of a sample under a phase contrast microscope. Because even after 2 days of culture the number of cells attached to the microcarriers was very low, this initiation method was not used.

In order to keep the cells in contact with the microcarriers for more time, the culture was done in two steps. First cells were allowed the to attach to the microcarriers in a plate with sporadic agitation, and second the cells and microcarriers were transferred to the bioreactor. Leaving the cells to attach over night ended in a strong attachment of cells to the microcarriers and formation of bridges between microcarriers. The clumps of cells-microcarriers were big and not stable enough for further manipulations, so that once they were put in the reactor formed masses of very different size. Actually the big size is not a problem, the problem is that the clumps became big under not controlled conditions. If this had taken place in a bioreactor where the fluidics is different there would be no problem with the size.

In order to avoid this, a lower inoculum and a bigger plate were used. $4 \times 10^5$ cells were put in a small petri dish and after 1 hour, the content was transferred to a D=10cm petri dish and left overnight. The day after, there were cells adhered to the carriers but still most of them had no cells. The content was transferred to the reactor at the slowest rotation rate, 8 rpm. After 72 hours of cell culture the number of cells on microcarries was of only 4 cells/microcarrier.

A bigger innoculum and a shorter period for attachment were tested. The cells were allowed to attach to the microcarriers for 2-4 hours only with sporadic agitation every 30 minutes for 1-2 minutes. The microcarriers and cells were transferred to the bioreactor and filled it with medium. The rotation rate was set to 8 rpm. After 48 hours not too many cells were on the microcarriers and even after 7 days cells did not cover the microcarriers.

Considering the above, it seemed that the cells needed more time to attached to the microcarriers, but they should not be static in order to avoid the formation of clumps. Then in order to increase the probability of attachment of the cells to the microcarriers after the resting period, the cells were transferred to the bioreactor and only half the of the reactor was filled with media. After 24 hours 1 of 10 microcarriers did not have cells, and the surface of the other microcarriers were covered in a 50-100% by cells. After 48 hours, the same was observed but more detached cells could be seen. As the cells grew and did not have surface to attach, they formed clumps of cells that detached form the surface of the microcarriers.

Then the initialization conditions were fixed. Considering that the number of cells that are attached to the microcarriers increases with time, in order to maintain them suspended, the rate of rotation has to be adjusted. This control was done daily, adjusting the rotation of the reactor, which varied from 8 to 20 rpm. Because of the high concentration of cells in the reactor, the media had to be changed in order to restore nutrients and to dilute toxic compounds that might be produced. Half the medium was substituted with fresh one (glucose concentration 1 g/l) every day. As figure 5 shows, the cell content of the reactor went form $10^6$ up to $10^7$ and declined after 120 hours of culture.

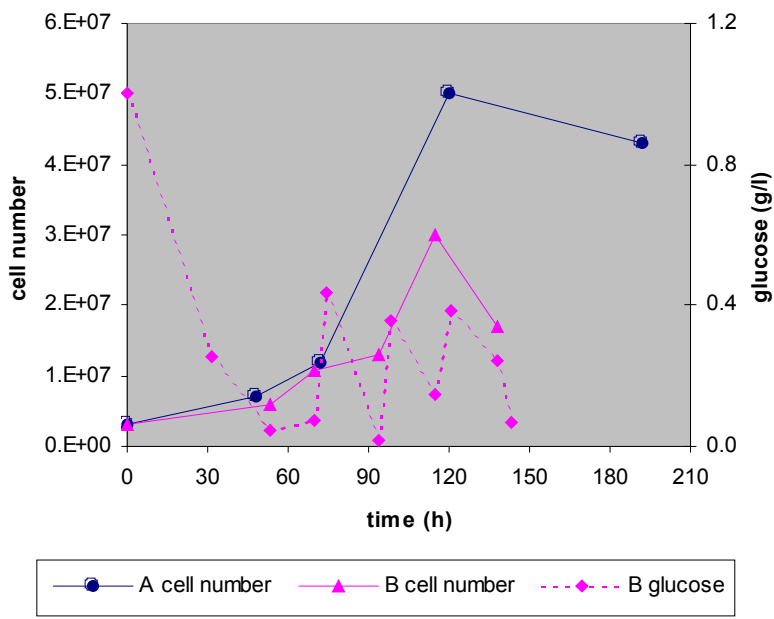

*Figure 5: HepG2 culture in RWV on Cytodex. Two different experiments A and B*

*Hepatocyte aggregates*

Half of the glucose was consumed in less than 30 hours so in order to permit the cell growth the media was changed daily. The frequency increased to twice a day after 60 hours of culture in order to avoid having concentration of glucose below 0,1 g/l, or not so often but using more concentrated glucose medium.

During the cell culture a qualitative study was done to evaluate the time that takes to cover the surface of the microcarriers, since after that point the growth rate may decrease. As it can be seen in figure 6 the microcarriers were covered almost completely after 72 hours. In spite of that, no decrease of the growth rate was observed between 72-120 hours of culture.

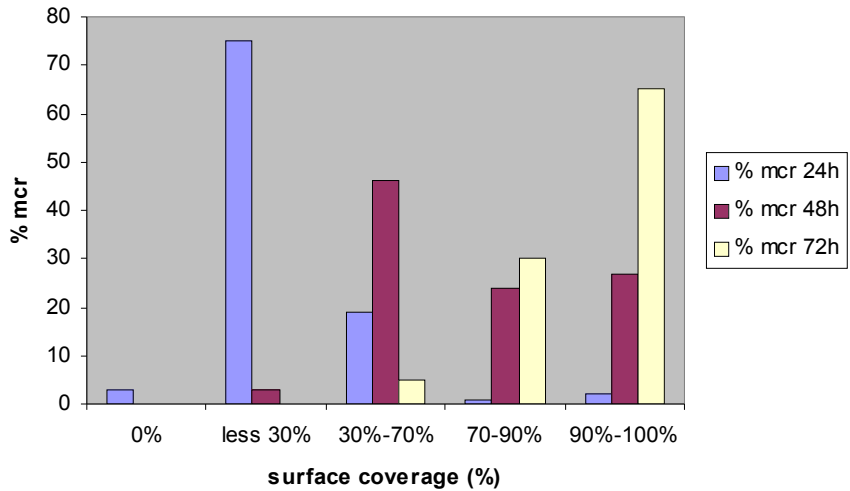

*Figure 6: Coverage of the surface of the microcarriers with cells*

### 4.2.2 Culture in 3D: spheroids

Other possibility of 3D cell culture is the formation of spheroids, which results form the self-aggregation of the cells among themselves. The cells after trypsinization are placed in the reactor filled until half of the volume. After 24 hours the reactor is completed with fresh media and the rotation speed adjust. As it can be seen in figure 7, the diameter of the spheroids increased during the culture, until no more than 80 µm.

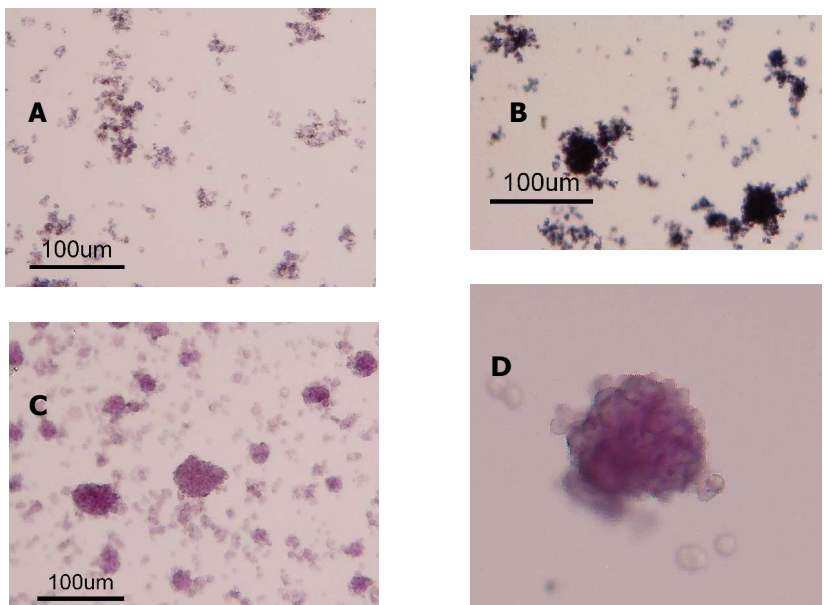

*Figure 7: Images of the spheroids obtained. (A) 24-hour culture; (B) 72-hour culture; (C) 96-hour culture; (D) Amplification of the image of one spheroid.*

During the time of the cultures, the cell number increased form $10^6$ to $10^7$ cells/ml. The glucose was monitored and the media change so that the concentration of glucose was not below 0.1 g/l. In order to avoid extremely low glucose concentration, a higher glucose concentrated media was used (4.5 g/l). An example of the growth curves obtained is shown in figure 8.

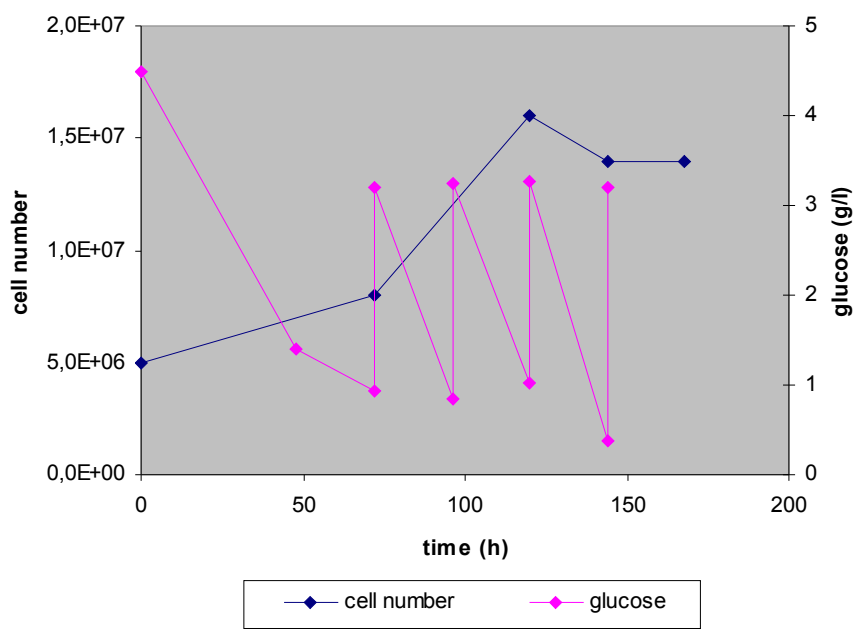

*Figure 8: Growth curve of HepG2 spheroids*

## *4.3 Toxicology tests*

In order to compare the response of the cells in 3D with the standard culture in 2D, toxicology test were carried out using standard 2 D culture and cell aggregates.

Toxicology test in 2D
As it can be seen in figure 9, the three parameters reported for the toxicological assays are % of DNA and MTT with respect to the control and glucose in the medium. The % of cells measured as DNA decreased drastically with the concentrations of APAP assayed. At 20 mM there is a reduction of about 60%, increasing up to 70% with higher APAP concentrations. The decreased in the viability measured as % of MTT seems to be a bit higher, which is expected, since the cells might be affected remaining alive but less viable.

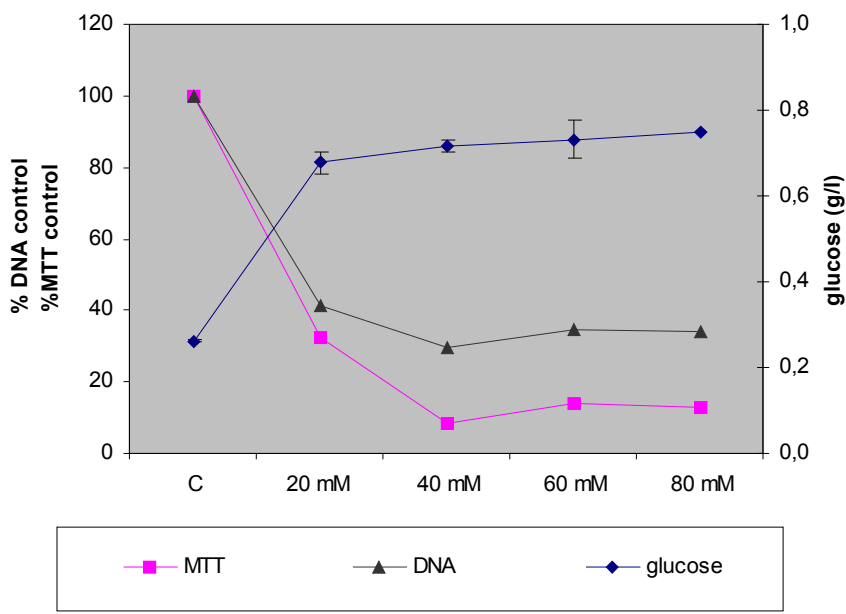

*Figure 9: 2D toxicological experiment*

The glucose remaining in the medium shows that the cells in the control consumed almost the 75% of the glucose of the medium, in contrast to the 30% of the 20mM to 25% of the 80 mM. This is in accordance with the fact that the amount of cells decreased with the amount of APAP added. The ratio of the glucose consumption and amount of DNA was calculated, to check if this relation could be related with the presence of the APAP; but no clear trend was found.

Even when not presented, LDH assays were carried out to estimate the amount of cell lysis. It seems that APAP interfered with the LDH assays. The values of the LDH content of the medium at t=0 h decreased just after the addition of APAP. That explained why the results of the assays at different concentrations of APAP after 24 hours of culture, were unexpectedly lower than the control.

## *4.4 Surface pattering*

See annex C

## *4.5 Other experiments performed*

*Characterization of the formation of aggregates*

During the cell aggregates formation, samples from cells and medium were analyzed. Protein and DNA assays were proposed to follow the increase of biomass. Since they requires only a sample volume of 50 µl and 150 µl respectively, their measurements could be done in triplicate to have a better estimation of the value. The results obtained were not reliable since the variation among the triplicates was very high. The sample preparation probably was one of the factors involved. Possibly the process of breaking the cells was not strong enough, as well as the breakage of the clumps that formed after defrosting the sample. LDH measurement of the medium was assayed to evaluate the degree of cell lysis in the reactor. The data obtained from these assays was difficult to interpret, since the overall value seemed to be very sensitive to uncontrolled factors. The fact that the medium had to be changed frequently, introduced a dilution factor that had to be considered. But in spite of that, sometimes the LDH measured after the change of the medium was higher than the one before and than the value before the following medium change. This probably was indicates that during the medium change, the system was being affected.

Because of that, the medium was changed carefully and rapidly, in order to avoid a long disturbance of the cell culture. In spite of that, the results did not improve.

*Toxicological experiments in 3D*

After carrying out the toxicological experiments using the cell aggregates, no convincing results were obtained. The range of concentration tested was form 5mM to 50mM. Total protein and amount of DNA were assay to measured the amount of biomass and the total number of cells and MTT as a viability assay. First, MTT assays did not give any acceptable results. Even though the value of absorbency decreased as the amount of toxicant increased, as it was expected, the fact that this has to be related with the amount of cell present in the assay, make this analysis inconclusive. Also some technical problems were present in the separation of the supernatant from the cells before the addition of the DMSO, and then may have affected the final result. On the other hand, the DNA and protein results were not consistent, probably presented the same problem as in the characterization of the aggregates.

In view of the above, when working with the cell aggregates, the total lysis of the cell was not being achieved. In order to improve the sample preparation, it is proposed to use a homogenizer buffer together with a physical treatment such as sonication or passage throw a thin syringe.

# 5. Discussion

In the 3 D culture systems, the production of aggregates of cells with and without microcarrier was achieved. At the beginning of the work, it was evident that the attachment to the microcarriers as well as the self-aggregation of the cells is very sensitive to current state of the cells after the subculture procedure. The cells have to be kindly treated during the trypsinitation, otherwise the aggregation is not attained in the suspension culture. Besides, it seems that the alterations of the culture regime have an impact on the state of the cells that cannot be avoided.

In the production of the cell aggregates, once the initialization conditions were established, the production of aggregates with Cytodex 3 and spheroids was done. In both cases, with an initial population of $10^6$ cell/ml in the reactor, the cells grew until $10^7$ cell/ml was after 72-96 hours of culture. The method used to follow the cell growth was direct counting of cells with trypan blue, since due to the complications of the system other methods were not successful. The information given by this method, complemented with image analysis that gave data of sizes and surface coverage, it is not enough. Since the three dimensional characteristic of the aggregates are the base of our system, other methods of characterization are required. The conventional methods employed results in an average response among the population of cells, so they do not permit to observe heterogeneity among cells in accordance with its localization in the aggregate.

The confocal microscope is an established tool in many fields of biomedical for imaging of cells within fluorescently-label tissues. The method of image formation in a confocal microscope is fundamentally different from that in a conventional wide-field microscope in which the entire specimen is bathed in light, and the image can be viewed directly by eye. In contrast, the illumination in a confocal microscope is achieved by scanning one or more focused beams of light, usually from a laser, across the specimen. The images produced by scanning the specimen in this way are called optical sections. It is then a non-invasive method, which uses light rather than physical means to section the specimen, enabling the automated collection of 3D data in the form of Z-series [28]. Using this technology is possible for example to have images of the distribution of alive and death cells in the cell aggregates, using two different fluorescent dyes that stain differently alive and dead cells. This would enable us to check the presence of necrotic center in the aggregates as well as the relation with the size of them.

This technology could also be applied in the toxicological test, giving information about the response of the cells to the toxic and also the spatial distribution of it in the aggregate. In this study, the number of cells and viability of them were the only response to the toxic measured. Other assays more specific to the response of the toxic are needed since comparison with standard culture is a goal.

Because we are interested in being able to check heterogeneity within the aggregates, the endpoints that we are interested in are those in which spatial distribution is involve.

The toxic used, APAP, is a hepatotoxicant bioactivated by specific P450s, mainly CYP 2E1, 1A2, and 3A4. One effective method of assessing cytochrome P450 isozyme activity is by the detection of resorufin formation from alkoxyresorufin substrates. Resorufin is a highly fluorescent compound that is easily measured. Alkoxyresorufin substrates have been used to probe the level of activity, inducibility, and substrate specificity of various P450 isozymes in rodent, and human hepatocytes as well as hepatic cell lines. [29]. The use of confocal microscopy permits the assessment of local P450 isozyme activity in three-dimensional tissue or tissue-like structure in situ by detection of local fluorescence intensity.

# 6. Conclusion

- A methodology for the formation of HepG2 cell aggregates with microcarriers and spheroids using a microgravity simulating bioreactor was developed. This involved:
    - Optimization of the initialization conditions of the culture
    - Identification of important factors affecting the formation of aggregates
    - Setting conditions of reactor operation: velocity of rotation and medium exchange
    - To establish protocols to monitor the cell culture

- A basic characterization of the cell aggregates was obtained. In order to visualized heterogeneity within the aggregates confocal microscopy will be used.

- Toxicology test has been carried out on cell aggregates. As a consequence of the results obtained, improvements in the methodology are needed. New experimental techniques, such as the assessment of cytochrome P450 isozyme activity will be used to determine the feasibility of the 3 D model (in comparison with the standard 2D culture).

- Microcontact printing was used successfully to produce a patterned surface. This surface was tested for BSA protein and the presence of the pattern confirmed by fluorescence microscopy and AFM. Immobilization of cell aggregates on protein-patterned surfaces will be attempt.

In the above conclusion is shown what remains to do in this first stage of the work proposed. Once ready to move to the next stage, the work will consist in the production of primary rat hepatocytes aggregates following the methods developed and it use in bioreactors with different configurations aiming to gain further understanding of the role of cell heterogeneity in the cooperative behavior of cells in vitro.

## 7. Acknowledgements

I would like to thank Dr. Petros Lenas and the Universitat Rovira y Virgili for the supervision and support. Also the Facultat de Medicina i Ciències de la Salut in Reus for giving me a place to work and specially the Biochemistry depart for their help. I also thank the Colloid and Interfaces group of the MPI and Dr. José Luis Toca for their contribution in the micro-contact- printing work.

## 8. Literature

1. Caplan, A.I., *Tissue engineering designs for the future: New logics, old molecules.* Tissue Engineering, 2000. **6**(1): p. 1-8.
2. Ferber, D., *Tissue engineering - Lab-grown organs begin to take shape.* Science, 1999. **284** (5413): p. 422-+.
3. Glicklis, R., et al., *Hepatocyte behavior within three-dimensional porous alginate scaffolds.* Biotechnology and Bioengineering, 2000. **67**(3): p. 344-353.
4. Benya, P., *Methods in cartilage research: Introduction and survey of techiques for chondrocyte culture.*, ed. M.A.a.K. K. 1990, London: Academic Press. 85-89.
5. Stone, A., *Investing in Tissue Engineering*, in *Business Week Online*. 1998.
6. Moore, S.K., *Tissue engineering pioneers fall to financial troubles.* IEEE Spectrum Online, 2004.
7. Allen, J.W., T. Hassanein, and S.N. Bhatia, *Advances in bioartificial liver devices.* Hepatology, 2001. **34**(3): p. 447-455.
8. Lee, W.M., *Medical Progress - Acute Liver-Failure (Vol 329, Pg 1862, 1993).* New England Journal of Medicine, 1994. **330**(8): p. 584-584.
9. Jauregui H.O., M.C.a.S.B., *Extracorporeal artificial liver support*, in *Principles of tissue engineering*, R.L.a.W.C. Rober Lanza, Editor. 1997, Landes Company.
10. Dixit, V. and G. Gitnick, *Artificial liver support: State of the art.* Scandinavian Journal of Gastroenterology, 1996. **31**: p. 101-114.
11. Salmeron, J.M., et al., *Bioartificial liver support for acute liver failure. First case treated in Spain.* Medicina Clinica, 2001. **117**(20): p. 781-784.
12. Rozga, J., et al., *Development of a Hybrid Bioartificial Liver.* Annals of Surgery, 1993. **217**(5): p. 502-511.
13. Watanabe, F.D., et al., *Clinical experience with a bioartificial liver in the treatment of severe liver failure - A phase I clinical trial.* Annals of Surgery, 1997. **225**(5): p. 484-491.
14. Stevens, A.C., et al., *An interim analysis of a phase II/III prospective randomized, multicenter, controlled trial of the HepatAssist (R) bioartificial liver support system for the treatment of fulminant hepatic failure.* Hepatology, 2001. **34**(4): p. 299A-299A.
15. Nose, Y. and H. Okubo, *Artificial organs versus regenerative medicine: Is it true?* Artificial Organs, 2003. **27**(9): p. 765-771.
16. NIH, *Reparative medicine: Growing tissues and organs*. 2001, National Institute of Health Bioengineering Consortium, BECOM.
17. Jungermann, K. and R.G. Thurman, *Hepatocyte Heterogeneity in the Metabolism of Carbohydrates.* Enzyme, 1992. **46**(1-3): p. 33-58.
18. Vortkamp, A., et al., *Regulation of rate of cartilage differentiation by Indian hedgehog and PTH-related protein.* Science, 1996. **273**(5275): p. 613-622.
19. Pipeleers, D., et al., *Physiological Relevance of Heterogeneity in the Pancreatic Beta-Cell Population.* Diabetologia, 1994. **37**: p. S57-S64.
20. Jungermann, K. and T. Kietzmann, *Role of oxygen in the zonation of carbohydrate metabolism and gene expression in liver.* Kidney International, 1997. **51**(2): p. 402-412.
21. Okubo, H., et al., *Novel method for faster formation of rat liver cell spheroids.* Artificial Organs, 2002. **26**(6): p. 497-505.
22. Unsworth, B.R. and P.I. Lelkes, *Growing tissues in microgravity.* Nature Medicine, 1998. **4**(8): p. 901-907.
23. Khaoustov, V.I., et al., *Induction of three-dimensional growth of human liver cells in simulated microgravity.* Hepatology, 1999. **30**(4): p. 508A-508A.
24. Wezel, A.L.v., *Growth of cell strains and primary cells on microcarriers in homogeneous cultures.* Nature, 1967. **216**: p. 64-65.

25. Avgerinos, G.C., et al., *Spin Filter Perfusion System for High-Density Cell-Culture - Production of Recombinant Urinary Type Plasminogen-Activator in Cho Cells.* Bio-Technology, 1990. **8**(1): p. 54-58.
26. Wilkening, S., F. Stahl, and A. Bader, *Comparison of primary human hepatocytes and hepatoma cell line HEPG2 with regard to their biotransformation properties.* Drug Metabolism and Disposition, 2003. **31**(8): p. 1035-1042.
27. Knowles, B.B., C.C. Howe, and D.P. Aden, *Human Hepatocellular-Carcinoma Cell-Lines Secrete the Major Plasma-Proteins and Hepatitis-B Surface-Antigen.* Science, 1980. **209**(4455): p. 497-499.
28. Paddock, S.W., *Methods in molecular biology: Confocal microscopy methods and protocols.* 1999, Totowa, New Jersey: Humana Press.
29. Wu, F.J., et al., *Enhanced cytochrome P450IA1 activity of self-assembled rat hepatocyte spheroids.* Cell Transplantation, 1999. **8**(3): p. 233-246.
30. Jungermann, K., *Zonation of Metabolism and Gene-Expression in Liver.* Histochemistry and Cell Biology, 1995. **103**(2): p. 81-91.
31. Christoffels, V.M., et al., *A mechanistic model for the development and maintenance of portocentral gradients in gene expression in the liver.* Hepatology, 1999. **29**(4): p. 1180-1192.
32. Tien, J. and C.S. Chen, *Patterning the cellular microenvironment.* Ieee Engineering in Medicine and Biology Magazine, 2002. **21**(1): p. 95-98.

# Annex A

*Glucose metabolism in liver*

In order to apply the new concept of designing, a liver tissue function had to be chosen. The glucose metabolism was selected because considerable information exists in liver physiology and liver biochemistry for the role of various enzymes involved. In addition considerable information exist for the substances in the hepatocyte microenvironments that induce the expression of these enzymes. No such detailed information exists for other liver metabolic functions.

According to the "metabolic zonation" theory, the enzymatic activities of metabolic functions in the liver, (oxidative and carbohydrate metabolism, amino acid and ammonia metabolism, cholesterol synthesis, xenobiotic metabolism, protective metabolism and plasma protein formation), are unequally distributed among the periportal, and perivenous, zones of liver lobules. The periportal hepatocytes specialize in glucose release and glycogen formation via gluconeogenesis and the perivenous hepatocytes specialize in glucose uptake for glycogen synthesis and glycolysis, (Fig. A1 and A2) [17]

The distribution of the metabolic activities in the periportal and perivenous hepatocytes allows the liver to operate as a "glucostat" regulating the blood glucose concentration at relatively constant levels, (80 to 120 mg/100ml). The perivenous hepatocyts removes glucose from the circulation during the absorptive phase after a carbohydrate-rich meal, and the periportal hepatocytes releases glucose during the postabsorptive phase when the glucose concentration in the blood is low in order to supply the central nervous system and the erythrocytes, (Fig. A1).

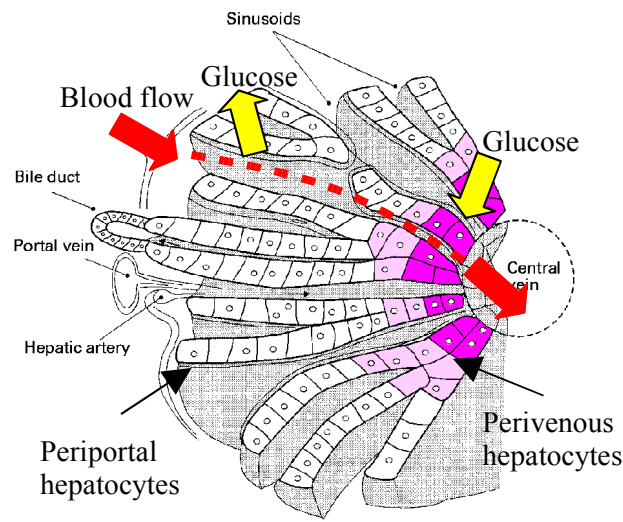

*Figure 10: Liver lobule. The liver lobule is composed of sinusoids, narrow veins for the blood flow. At the walls of the sinusoids are the hepatocytes. Periportal hepatocytes are situated close to the blood entrance and preferentially release glucose, while perivenous hepatocytes at the blood exit consume glucose*

The metabolic reactions of the glucose metabolish and the enzymes that participate are included in the fig. A2. In few words the glucose that is taken up during the absorptive phase is incorporated into glycogen and then when the glycogen stores are replenished it is degraded to lactate. Lactate then is transported by the circulation to the periportal cells where it is converted via gluconeogenesis to glycogen. Glycogen is degraded to glucose in the periportal cells during the post-absorptive phase. Then, besides the heterogeneity in the metabolic functions of hepatocytes, another critical factor is the metabolic cooperativity through the transport of lactate.

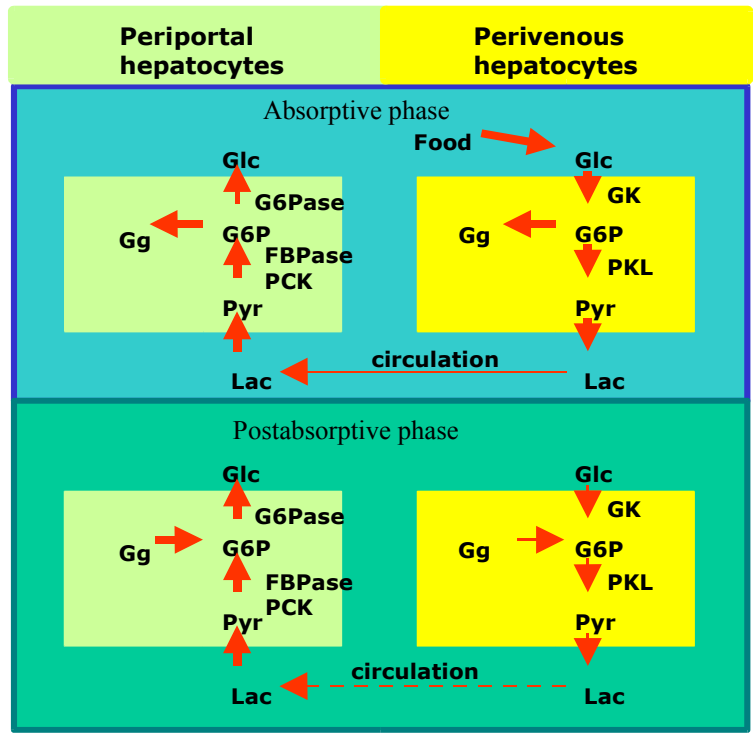

*Figure 11: Glucose homeostasis in the liver. Glc, glucose; Gg, glycogen; Lac, lactate; Pyr, pyruvate; G6P, glucose 6-phosphate; G6Pase, glucose 6-phosphatase; GK, glucokinase; PKL, pyruvate kinase type L; FBPase, fructose 1,6-biphosphatase; PCK, phosphoenolpyruvate carboxykinase [30]*

Most of the enzymes of carbohydrate metabolism have higher activity in the periportal or the perivenous hepatocytes. The key enzymes of gluconeogenesis, glucose-6-phosphatase (G6Pase), fructose-1,6-bisphosphatase (FBPase), and phosphoenolpyruvate carboxykinase (PCK), are periportal, and the key enzymes of glycolysis, glucokinase (GK), and pyruvate kinase isoenzymes L (PKL), are perivenous. The unequal spatial distribution of enzyme activities is due to either regulation at pre-translational level (e.g. phosphoenolpyruvate carboxykinase) or regulation at translational or post-translational level (e.g. glucokinase, pyruvate kinase L),[17]).

The zonal patterns of the gene expression related with the glucose metabolism have been categorized as "gradient" type because all the hepatocytes are able to express these genes but the level of expression depends on the position of the cell along the sinusoid. The gradients of substances arising from the exchange of metabolites between the hepatocytes with the flowing blood induce differential gene expression [31]. The gradients related with the glucose metabolism are the gradients of oxygen and the hormones glucagon and insulin. Most important is the gradient of oxygen which is the steepest falling by 50% from the periportal to the perivenous zone.

# Anex B

## *Protocols*

*Cell defrost*

Cells were kept in $N_2(l)$ until use. To start the culture, one ampoule was defrosted by addition of warm DMEM (37ºC). DMEM was slowly added to the cryotube and was transferred to two 15 ml tube. Once completely defrost, it was centrifuged at 1000 rpm for 3 minutes. The supernatant was discarded and more DMEM was added to completely eliminate the DMSO. It was again centrifuge for 3 minutes at 1000 rpm and the medium discarded. Finally cells were resuspended with complete medium and transferred a to tissue culture flask for culturing.

*Subculture*

Cells in flasks were observed under a phase contrast microscope on alternating days to inspect the state of the cells. After confluence is reached, which means that all the available growth area is utilized and the cells make close contact with one another, cells will overgrow, which means that probably if subculture a longer lag phase will be found. Then, subculture is done at 80% confluence.
Cells were twice rinsed with phosphate buffer solution (PBS) to remove the cell culture media from the flask. The PBS was removed and trypsin formed by a 1:10 dilution of 10X trypsin in PBS, was added to the flask which was incubated for 1 minutes at 37 °C. After that, trypsin was agitated by a pipette to facilitate cell detachment from the plate. FBS was added to stop the action of trypsin The cell suspension was removed from the flask and placed in a sterile 15 mL tube and centrifuged for 3 minutes at 1000 rpm. The supernatant was removed and replaced with complete media, and the cells were resuspended. These cells could either be replated on tissue culture flasks to expand the culture or used for the different experiments.

*Sampling 2D culture*

First a sample of the medium was directly taken from the plate. After 3 minute-centrifugation at 1000 rpm, they were stored at –80ºC until assayed. The rest of the medium was discarded and the cells washed with twice with PBS. The PBS was removed and trypsin formed by a 1:10 dilution of 10X trypsin in PBS, was added to the flask which was incubated for 1 minutes at 37 °C. After that, trypsin was agitated by a pipette to facilitate cell detachment from the plate. FBS was added to stop the action of trypsin. The cell suspension was removed from the plate and placed in an eppendorf tube and centrifuged for 3 minutes at 1000 rpm. The supernatant was removed and cells resuspended in 1 ml of PBS. In order to eliminate any rest of FBS, it was again centrifuged idem before. Cells resuspended in 1 ml of PBS were kept at –80ºC until assayed.

### 3-D cell culture

*Replacing culture media*

When Cytodex was used the content of the reactor was allowed to sediment for 15 minutes and then the media was changes. Because the spheroids did not sediment easily, media was extracted with a syringe through one of the port through a 0,5 mm filter in order to avoid the lost of cells. Afterwards pre-warmed complete medium was added, and absence of bubbles checked.

*Sampling 3D culture*

1 ml sample of the content of the reactor was taken while it was rotating with a syringe, and the same volume of medium was added in order to avoid bubbles inside the reactor. Cells and medium were separated for later assessment. The medium was store at –80ºC until assayed, and the cell sample taken as explained in cell harvesting section.

Cell harvesting

The spheroids as well as the Cytodex 3 covered with cells could be visualized directly by phase contrast microscopy. In order to be able obtain a homogenous cell sample, trypsinition of the cells was done in order to detach cells from the microcarrier and/or from other cells.

For cells on Cytodex 3, the microcarriers were allowed to settle, culture medium was removed, and the microcarriers were washed for 5 minutes in PBS containing 0,02% (w/v) EDTA, pH 7,6. The EDTA-PBS is removed an replaced by trypsin-EDTA and incubated with occasional agitation. After 15 min. the action of the trypsin is stopped by addition of FBS. The sample is centrifuge again and suspended in 1 ml of PBS.

In the case of the spheroids, they were centrifuged for 5 minutes at 1000 rpm. Culture medium was removed and washed for 5 minutes in PBS containing 0,02% (w/v) EDTA, pH 7,6. After 3 minutes centiguation at 1000 rpm, the EDTA-PBS was removed an replaced by trypsin-EDTA and incubated for 15 minutes with occasional agitation. The action of the trypsin was stopped by addition of FBS. The sample is centrifuge and suspended in 1 ml of PBS.

## *Measurements*

*Direct cell count with trypan blue*

Typan blue is a dye that is taken up by dead cells, but excluded by living cells. 500 μL of a very well mixed cell suspension obtained according cell harvesting protocol, was place in an eppendorf and added equal volume of the diluted dye. Cell count in a hemacytometer was done.

*DNA with Hoechst*
100 μL of the cell sample was placed in a 96 well plate by triplicate and 50 μl of Hoechst solution is added in order to obtain a final concentration of 10 μg/ml. After 1-hour incubation at 37ºC, fluorescence is measured.

*Viability with MTT*
After cell harvesting, the cells were resuspended in complete medium, and 100 μL cell sample was placed in a eppendorf tube by triplicate. 10 μL of 5mg/ml MTT solution is added to each tube. After 3 hours of incubation at 37ºC, and 3 minutes centrifugation at 1000 rpm, the medium is discarded and 100 μL of DMSO is added. To measure the absorbance, the sample was transferred to a 96 well plate and absorbance read at 550/620 nm with BioWhittaker Kinetic QCL plate-spectrophotometer.

# Annex C

## *Surface pattering using micro-contact printing*

Chemically microstructured surfaces have gained increasing attention in the cell biological/biomaterial community in the past few years. There is a broad field of practical applications of micro-structured substrates, ranging from tissue engineering through cell array-based biosensors to artificially designed neuronal networks. Among the numerous methods to pattern biomolecules on a solid surface, microcontact printing (μCP) is becoming increasingly popular. There are several reasons for this popularity, μCP is a simple and cost-effective method, does neither require clean room instrumentation nor absolutely flat surfaces, plus it offers a way to create complex patterns (yet with some geometrical constraints) on surfaces. Besides, in contrast to patterning methods that are based on photolithography, μCP does not subject any of the patterning materials to harsh chemical or physical treatments that most biological molecules. All printing occurs in ambient conditions, and all post-printing steps occur in aqueous solutions that do not denature proteins. In addition, the use of a soft deformable stamp as the patterning element allows patterining over large areas [32].

The objective of this work was to learn the microcontact technique, apply it to obtain a protein-patterned surface of bovine serum albumin (BSA) and use fluorescence microscopy and atomic force microscopy (AFM) to confirm the presence of the pattern.

## *Materials and methods*

The silicon masters used for producing the microcontact-printing were provided by GemSin, Germany. Poly(sodium-4-styrensulfonate) (PSS, Mw=70000 g/mol), poly(allyamine-hidrochloride) (PAH, Mw=700000 g/mol), poly(ethyleneimine) (PEI, Mw= 25000 g/mol) were purchased form Aldrich. Poli-L-lysine grafted polyethilen glicol (PLL-g-PEG/PEGbio (20)-[3.5]-(2,3,4) (50% bio) was provided by Surface Solutions. It is a graft copolymer with a PLL backbone of molecular weight of 20 kDa, a grafting ratio of lysine-mer/PEG side chain of 3.5 with molecular weight of 2 kDa. Rhodamine-B-isothiocyanete label PAH (RBITC-PAH) was prepared as described by Richter *et al* and Ibarz *et al*. The FITC-PAH was purchased from Capsulution NanoScience AG(Berlin, Germany). BSA form Sigma was labeled with TRICT using standard methods.

PDMS base silicon elastomer Sylgard 184 and the curing agent (Dow Corning Midland, MI) were mixed in a 10:1 ratio, and degassed in a vacuum. The masters were covered with the pre-polymer, degassed again, and cured for 12h at 60ºC. The stamps were cut out and hydrophilized in an air-plasma cleaner (Harrick PDC-32G-2, pressure: 5 mbar, 1 min). Before the stamping was carried out, a 1mg/ml solution of the polyelectrolyte to stamp (PAH in 500 mM NaCl or PLL-g-PEG) was spread on the stamp. After 15 minutes, the stamps were briefly rinsed with water an excess solution was removed in a steam of nitrogen. The inked stamps were brought into contact with clean and dried coverslips with or without coating for 20 minutes and then removed. The structured surfaces were carefully rinsed with water. The adsorption of the BSA was done by covering the area of interest with a drop of the solution and left for 5 minutes and afterwards carefully cleaned with water. The cleaning of the coverslips prior coating with polyelectrolytes was based in the RCA method. Briefly, the coverslips were put in 98% ethanol and sonicated for 15 minutes. After rinsing with water, they are immersed in a solution of $H_2O:NH_4OH:H_2O_2$ 5:1:1 at 70ºC for 15 minutes. After intensive rinsing with water, the coverslips are ready to use. In the case of PLL-g-PEG stamping after 15 minutes sonication in ethanol 98% were followed with 2 minutes of plasma cleaner.

Fluorescence microscopy imaging was carried out on Zeiss Axiovert 200 (Zeiss, Germany) with a Axiovert LDA-Plan 10X, light source was Hg vapor lamp. A Zeiss Axiocam HR high-resolution monochromatic camera was used. Atomic force microscopy (AFM) measurements were done with a Nanoscope IIIa AFM in tapping mode.

## Results and discussion

1) Polyelectrolyte patterning

The first µCP attempts were done using fluorescently label PAH in order to be able to check easily the stamped pattern. The RBITC-PAH pattering on glass coated with one layer of PEI and PSS was obtained as it is shown in fig.C1.

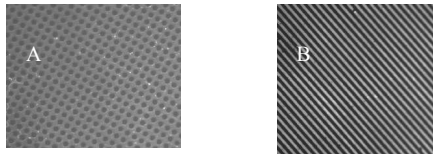

*Figure 12: PAH patterns on PEI-PSS coating on glass coverslips. (A) PAH is the complementary area of D= 10µm circles; (B) strips of aprox. 5µm.*

Then it was of interest to check the possibility of selectively covering the surface of a patterned substrate with a polyelectrolyte. The selective adsorption of FITC- PAH was tested. As it can be seen in the fig. 2, the FITC-PAH/RBITC-PAH pattern is observed clearly. Applying the same technique, PLL-g-PEG was stamped and afterwards RBITC-PAH was added. Figure C2 shows that even though it is not so clean, the pattern polyelectrolyte pattern was obtained.

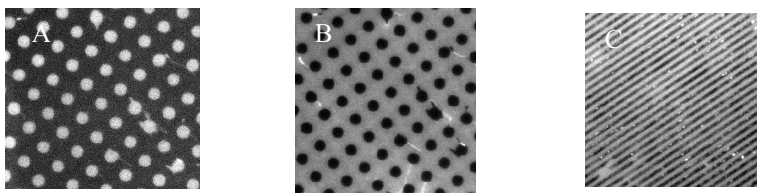

*Figure 13: Pattern of FITC-PAH and RBITC-PAH. (A) Image of the FITC-PAH pattern; (B) Image of the RBITC-PAH pattern; (C) PLL-g-PEG pattern and RBITC-PAH.*

2) BSA patterning

In order to obtain the resistant protein pattern PLL-g-PEG was used. After stamping the glass slide, TRICT-BSA was put on the surface of the slide. Fig. C 3 and C4 proves that the protein resistant surface of the polyelectrolyte prevents the BSA to attach in those areas where PLL-g-PEG was stamped.

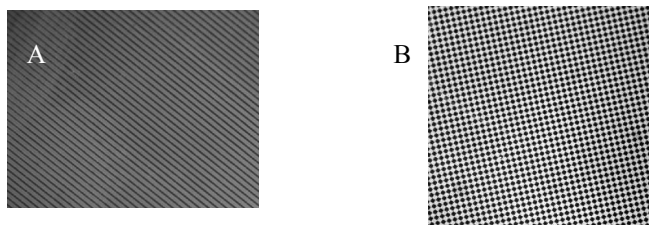

*Figure 14:TRICT-BSA on glass with protein resistant pattern of PLL-g-PEG. (A) strips 10 µm width, (B) squares .*

AFM measurements of the high differences of the surface were done by AFM before and after the addition of the BSA. Figure C4 shows one experiment of BSA adsorption.

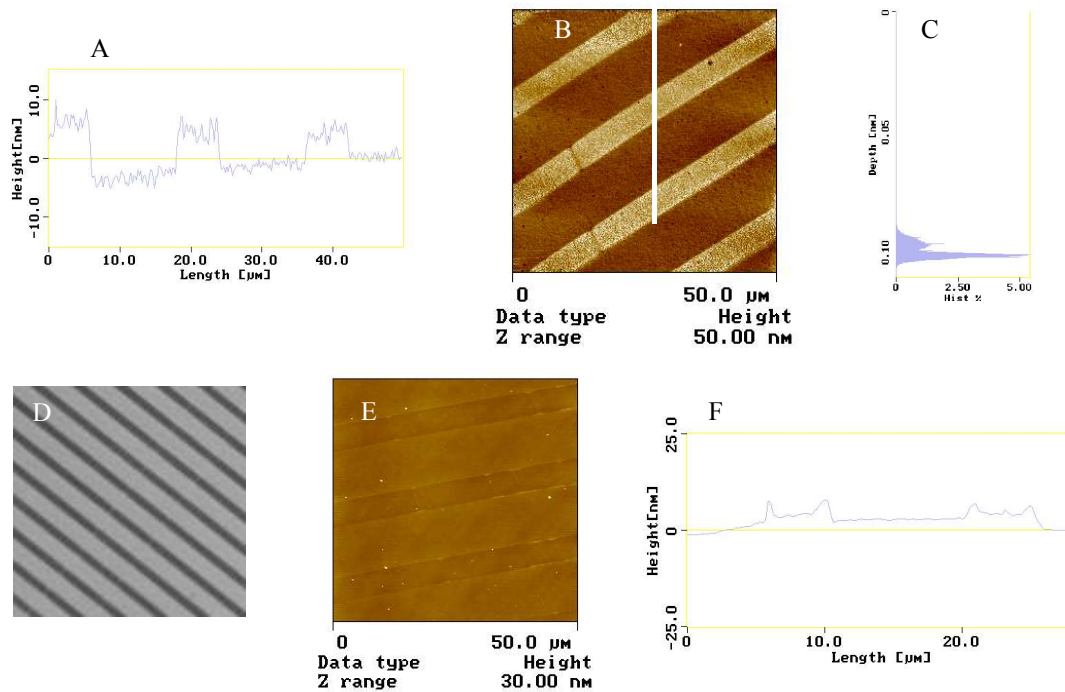

*Figure C4: PLL-g-PEG pattern on glass surface before the protein adsorption. (A) High differences of the section shown in figure B before any mathematical treatment. (B) Image of the pattern after performing a second order flattening. (C) Result of the bearing analysis of data of figure B. (D) Fluorescence microscope image of the patterning after the adsorption of BSA. (E) AFM image of the patterning after the adsorption of BSA. (F) Section average of the surface obtained after the adsorption of the BSA.*

As it can be seen in figures C4, it was possible to confirm the presence of the PLL-g-PEG. The heights differences measured were in two sets of experiments of 2,3±0,3 nm (n=3) and 9,3±2,5 nm (n=13). As it can be seen it was no possible to obtain a reproducible height difference in the stamping.

After the adsorption of the BSA, the differences in height were reduced, in the first case to 1,6±0,4 nm and in the second case 1,5±1,0 nm.

In figure C3, it is clearly seen that the broader strips that in figure C4 A and C4B correspond to the bare glass are covered with BSA as shows figure C4D, showing that the BSA has selectively adsorbed in that area.

In both cases the amount of protein adsorbed was such that after the adsorption, the surface obtain had almost no difference in height, then the amount of protein adsorbed seems to have been controlled by the height of the PLL-g-PEG pattern, since in both cases the time and concentration of the protein was the same.

*Conclusion*

A BSA patterned surface was attained by μCP of PLL-g-PEG and it presence checked with AFM. Further work will involve choosing a protein of interest, pattern it using this technique, and afterward test if cell aggregates can be selectively immobilized on the surface.